\documentclass[sigconf]{acmart}

\usepackage{enumitem}
\usepackage{soul}

\AtBeginDocument{%
  \providecommand\BibTeX{{%
    \normalfont B\kern-0.5em{\scshape i\kern-0.25em b}\kern-0.8em\TeX}}}

\acmPrice{15.00}
\acmISBN{978-1-4503-XXXX-X/18/06}

\begin{document}

\title{Towards Activity-Centric Access Control\\ for Smart Collaborative Ecosystems}



\author{Maanak Gupta}
\affiliation{%
  \institution{Department of Computer Science, \\ Tennessee Technological University}
  \city{Cookeville, TN}
  \country{USA}
}
\email{mgupta@tntech.edu}

\author{Ravi Sandhu}
\affiliation{%
  \institution{Institute for Cyber Security and NSF C-SPECC Center,  \\ Department of Computer Science}
  \city{University of Texas at San Antonio}
  \state{TX}
  \country{USA}
}
\email{ravi.sandhu@utsa.edu}
\renewcommand{\shortauthors}{Gupta and Sandhu.}

\begin{abstract}
The ubiquitous presence of smart devices along with advancements in connectivity coupled with the elastic capabilities of cloud and edge systems have nurtured and revolutionized smart ecosystems. Intelligent, integrated cyber-physical systems offer increased productivity, safety, efficiency, speed and support for data driven applications beyond imagination just a decade ago. Since several connected devices work together as a coordinated unit to ensure efficiency and automation, the individual operations they perform are often reliant on each other. Therefore, it is important to control what functions or activities different devices can perform at a particular moment of time, and how they are related to each other. It is also important to consider additional factors such as conditions, obligation or mutability of activities, which are critical in deciding whether or not a device can perform a requested activity. In this paper, we take an initial step to propose and discuss the concept of Activity-Centric Access Control (ACAC)  for smart and connected ecosystem. We discuss the notion of \textit{activity} with respect to the collaborative and distributed yet integrated systems and identify the different entities involved along with the important factors to make an activity control decision. We outline a preliminary approach for defining activity control expressions which can be applied to different smart objects in the system. The main goal of this paper
is to present the vision and need for the activity-centric approach for access control in connected smart systems, and foster discussion on the identified future research agenda.

\end{abstract}



\maketitle

\section{Introduction}


Cyber Physical Systems (CPS) and connected technologies are transforming how humans interact with engineered systems. These sensors and Internet of Things (IoT) based systems are built and depend upon the seamless integration of physical and computational components along with the emerging technologies of Artificial Intelligence (AI) and Machine Learning (ML) to support fully automated data driven ecosystems. Innovative smart CPS applications in the domain of energy, healthcare, transportation, building, agriculture, aeronautics and medicine, have offered convenience, comfort and efficiency for application developers and end users. As such according to the United States National Science Foundation, \textit{a “smart and connected community” is, in turn, defined as a community that synergistically integrates intelligent technologies with the natural and built environments, including infrastructure, to improve the social, economic, and environmental well-being of those who live, work, learn, or travel within it.} Everyday life is becoming increasingly dependent on these systems, often with dramatic improvements.

Ensuring the trustworthiness, security and privacy of information is a formidable challenge in regard to both technical and policy aspects. Deployment of secure mechanisms for critical infrastructures and CPS is important to assure resilience and protection from adversaries including state supported entities and foreign based actors. While cybersecurity is a strong national priority and much progress has been made to ensure protection from cyber-attacks, CPS security raises a host of new challenges. The convergence of physical and cyber world has introduced new attack modes which are automated, hard to analyze and engender substantial risk in maintaining the integrity of physical as well and cyber resources. Important challenges to secure CPS and IoT include threat modeling, proposing mathematically grounded fundamental security approaches along with continuous vulnerability assessment, and designing adaptable autonomous defense mechanisms to thwart rapidly evolving cyber and physical threats in this growing, connected, collaborative and distributed ecosystem.


Access control mechanisms have been extensively used to limit unauthorized access to resources and secure communication among objects. Several cloud and edge assisted IoT and CPS architectures have been proposed supporting traditional access control models such as Discretionary (DAC) \cite{sandhu1994access}, Mandatory (MAC) and  Role Based Access Control (RBAC) \cite{sandhu1998role}, as well as more fine grained such as Attribute-based access control (ABAC) \cite{jin2012unified, gupta2016mathrm} offering flexibility in distributed and dynamic environments. However, traditional access control models are not well suited to address access control requirements in distributed systems where resources are managed by different entities, and are connected via single or multi cloud platforms. Such environments and applications will proliferate as we move to more data driven and automated world. We believe it is necessary to adopt a different viewpoint to develop access control approaches for such task and activity driven connected systems which work collaboratively to fully automate the entire ecosystem.


 



\subsection{Motivation for Activity-Centric Access Control Systems}

In a connected ecosystem, different operations and workflows in a coordinated unit are usually interdependent and collaborative. An activity is a unit of work which is performed by a device, reflecting its current state. For example, consider an aerial drone performing \textit{digital imaging} or \textit{soil health scanning} of a large farm. Here \textit{digital imaging} and \textit{soil health scanning} are the \textbf{activities} which are initiated by a subject (which can be a user or another smart device) via an operation. Activity is considered the fundamental operation in our proposed approach. Some activities can be related to each other and consequently should be done in a particular order such as before, concurrent or after other activities. Some activities may only be initiated by certain users or devices, or can only be done during a particular time slot, or there may be a threshold on an activity's occurrence (i.e. how many times or at what rate the activity is allowed). Therefore, such relation and contextual factors among different entities involved in requesting an activity must be considered to decide if the activity is allowed or denied in the system at a particular moment of time. The current proposed access control models including discretionary, role based or attribute based solutions proposed for IoT and CPS connected ecosystem do not capture the notion of activity, i.e the current state of the devices in the system, to control a new requested operation which will result in activities performed by different co-related devices. 

The main contributions of this paper are as follows.
\begin{itemize}
    \item We highlight key distinctions of the proposed activity-centric access control approach for collaborative systems comparing it with related but fundamentally distinct access control systems designed for enterprise and social computing.
    \item We propose the notion of an \textit{activity} with respect to the connected ecosystem and define what are the different abstractions involved along with the contextual factors important to make an activity control decision.
    \item We provide an activity-centric access control framework by identifying different relationships among various activities in a connected collaborative ecosystem.
    \item We also offer our preliminary thoughts and an approach in defining an activity control expression which can be applied to different smart objects in the system. 
    \item We highlight future research needs and outline an agenda to fully develop, mature and use the ACAC security models in smart collaborative ecosystems.
\end{itemize}
 
The rest of the paper is organized as follows. Section 2 discusses background with respect to the prior related models proposed in different domains, and explains how our proposed approach is fundamentally different from these prior approaches. Section 3 defines activity primitives and relationship characterization among different activities in a connected and collaborative ecosystem, along with factors impacting an activity-centric access control decision. Section 4 provides an approach to specify the proposed activity control expression, followed by a future research agenda in Section 5. Section 6 summarizes and concludes our work.


\section{Background and Related Work}
In this section, we discuss some prior related models and explain how our approach is different from these.
\subsection{Prior Related but Distinct Models}

\subsubsection*{Task-Based Authorization Control (TBAC)}
The TBAC model \cite{thomas1998task} enables ``active security'' concept, which offers the abstractions and mechanisms for the active runtime management of security, as tasks progress to completion especially relevant in workflows and transactions environments. Here the permissions are constantly monitored and activated-deactivated in accordance with emerging context associated with progressing tasks, providing tighter just-in-time need to do permissions. TBAC supports the fundamental abstraction of an \textit{authorization step} which is a single act of granting permissions (similar to a signature in the paper forms by an individual), referred to as \textit{enabled-permissions}. Further, these permissions are good only for a limited period of time, and an associated period of validity and lifecycle is associated with every authorization step. When a usage count reaches its limit, the associated permission is deactivated and the corresponding action is no longer allowed. Each authorization step corresponds to some activity or task within the broader workflow context.

\subsubsection*{Usage Control (UCON)} The UCON model \cite{park2002towards,sandhu2003usage,ps04,phb06} covers obligations, conditions, continuity (ongoing controls) and mutability of attributes, to determine an access control decision. Obligations require some action by the subject so as to achieve or sustain access, such as clicking accept on the terms of conditions of a license agreement. Conditions cover the environmental factors that predicate access such as the time of the day or risk level. In addition, continuity reflects continuous enforcement of access control is done by evaluating usage requirements throughout usages. The model also supports change in the attributes of the users and objects as side-effects of subject's actions. In case of attributes mutability, updates are supported before (pre), during (ongoing) or after (post) usages. Access control decision-making is done either before (pre) or during (ongoing) exercise of the requested right. The  UCON model supports applications such as trust management, digital rights management and privacy protection within a unified framework.

\subsubsection*{Activity-Centric Access Control (ACON)} The ACON model\footnote{Both ACON and our new model ACAC are acronyms for Activity-Centric Access Control. Given their motivations and intended application domains we believe this reuse of the underlying term will not cause confusion. Where necessary the acronyms can be used to disambiguate.}  \cite{park2011acon,psc11} is motivated by social computing systems (SCSs) like Facebook, in which a user performs activities not only on shared content but also against target users (a user pokes another or recommends friends). Furthermore, in SCS, there are activities performed by the system to provide services and resources that can promote user interactions or sustain the SCS provider’s business. These SCS’s activities also need to be evaluated for control decision since users may not want their shared information to be used for SCS’s analysis or may not want to receive some of these SCS services. From access control point of view, both users’ control activities and systems’ automated activities are rather unique to SCSs and seldom considered in traditional access control models. This work proposes activity as a key concept for access control in SCSs. ACON supports personalized user privacy control by utilizing individualized user policies/attributes and
resource policies/attributes. 

\subsection{How Activity-Centric Access Control (ACAC) is Different?}

Our proposed approach for activity-centric access control offers distinct but converging synergy with the above mentioned models, offering run-time access control considering the context, usage and various activities occurring (or have occurred) in the broader collaborative and connected smart ecosystem context. The core concept of \textit{Activity} is central to our proposed framework and is natural for smart devices. An activity on a device is initiated due to an operation from a subject (a device or a user), performing a short-lived or prolonged function (aka task). In a collaborative ecosystem such tasks (or activities) will typically be related with each other. For instance, in smart farming or smart manufacturing domains one activity may lead to another related activity, or the occurrence of an activity may restrict the initiation of another activity. As an example, water sprinkler cannot be turned ON while weed spraying is taking place. To our understanding no prior research has provided a perspective to support access control in IoT and CPS collaborative environment with \textit{activity} as the central notion. 

With respect to TBAC, our approach considers the relation among different activities occurring in the system to determine an access decision to perform a new activity requested by the subject. This condition is checked in addition to the permission for the subject to perform an activity. This is distinct from TBAC model, which does not consider relationship among other concurrent activities, but rather is focused on workflow dependencies amongst tasks. Further, TBAC only considers one object to have run-time control of permissions, while in ACAC multiple dependent activities can occur on different objects. UCON offers obligations from the users (subjects) and environmental conditions, which is different than the \textit{pre} or \textit{post} obligations of activities as required in our proposed approach. In addition, this obligation can be from the same subject, or different users/devices in the system. For example, to start an activity A from Tom on object Ob1, Bob has to start activity C on object Ob2. Similarly, conditions can be beyond just the environmental or system factors, and may also include other activities or the relationship among user/devices which have initiated different activities in the system. Limit controls on activities are supported, such as where a user can perform certain activity only \textit{twice} a day or an activity can be allowed only \textit{twice} irrespective of the initiating subject. The ACON model was designed taking into consideration social computing systems. However, it does not consider the current activities in the system to limit new activities. Also, the activities in the SCS are not persistent for a length of time, and are more similar to short-lived actions. In our proposed framework, an operation will result in an activity which will sustain for a longer duration of time, for example, water spraying will done for 1 hour. The activity-centric framework for smart connected ecosystem is fundamentally different from the related and established using similar notations.


\subsection{Access Control Requirements in Smart and Connected Ecosystems}

The goal of smart and connected ecosystems is to bring new levels of economic opportunity and growth, safety and security, health and wellness, accessibility and inclusivity, and overall quality of life, by offering data driven applications to end users. Further, these technologies are
combined with elements of the physical world (e.g., machines,
devices, structures) to create smart and intelligent systems that offer increased effectiveness, productivity, and speed. These connected yet distributed systems are supporting services in domains including manufacturing, energy, transportation \cite{gupta2018authorization, gupta2019dynamic}, medical, city, building, and agriculture \cite{sontowski2020cyber, gupta2020security}, offering data driven and intelligent AI driven efficient environments. With emergence of fully automated manufacturing and CPS domains, trustworthiness and security when considering a action related to AI and operations in the system is critical. Different cloud service providers, including Amazon Web Services,\footnote{https://aws.amazon.com/iot/} Google Cloud IoT,\footnote{https://cloud.google.com/iot-core} and Microsoft Azure\footnote{https://azure.microsoft.com/en-us/overview/iot/}, have dedicated IoT and CPS platform catering to diverse applications and use-cases supporting both cloud and real-time edge-based user applications and services.

Access control mechanisms are essential to secure data and resources in a connected shared ecosystem. These models and mechanisms offer solutions to restrict which smart device can be controlled by other resources, which can share data and with whom, what applications can gather the data from on-field devices, cloud communication and data exchange etc. Controlling which users applications or devices can operate or access other connected devices, get data from other devices, securing data \cite{gupta2018attribute,gupta2017object} in the cloud or local edge gateway and also in the transit are important concerns that need to be addressed. This problem magnifies when data and resources are distributed \cite{hu2018access} and spread across different entities administered by different units. Several approaches have been used in enforcing access control policies, including cryptographic mechanisms \cite{bertino2019iot}, capabilities, access control lists, and policy based solutions. Access control reference monitor will allow operations only if it has a policy that grants permission for requested operations. Attribute-based access control (ABAC) \cite{gupta2016mathrm,jin2012unified, gupta2021reachability} supports fine
grained authorization capabilities for resources offering flexibility in a distributed multi-entity environment where the attributes of entities along with contextual information are used for access and communication authorization decisions. Several cloud service platforms including AWS and Google Cloud provide policy based \cite{bhatt2017access,gupta2020access} security solutions to control among different smart entities and applications in the connected ecosystem.

In the past, numerous access control models \cite{alshehri2016access, bouij2015smartorbac, tian2017smartauth, ye2014efficient, gupta2020secure} and mechanisms \cite{schuster2018situational, jia2017contexlot} have been proposed to address authorization needs in both edge and cloud assisted IoT and CPS architectures. Ouaddah et al \cite{ouaddah2017access} presented a comprehensive review of IoT access control models, whereas survey based studies \cite{he2018rethinking} in smart home IoT have also highlighted the need for novel perspective of access control based on the relationship among the device owner and the subject. Work by Fern{\'a}ndez et al \cite{fernandez2020data} proposed a novel data collection and sharing model for cloud-IoT architectures providing plug-in module to support IoT application development. 
Convergent Access Control framework was recently proposed by Bhatt and Sandhu \cite{bhattconvergent} which highlights the need for synergistic convergence of access control models at both  policy and enforcement layers which can address the evolving access control requirements of dynamic applications for future smart communities.   



\section{Activity-Centric Access Control (ACAC) Framework}

In this section we describe a preliminary ACAC framework within which specific formal models, policy languages and enforcement architectures can  subsequently be developed. Our approach is to motivate various components of this framework by identifying interesting use-cases. We first discuss different entities which will be relevant in activity-centric models, and then discuss the identified characteristics reflecting relationship among activities in the connected collaborative ecosystem. 
 

\subsection{Activity Primitives}

\begin{figure}[t!]
\centering
\includegraphics[width=\columnwidth]{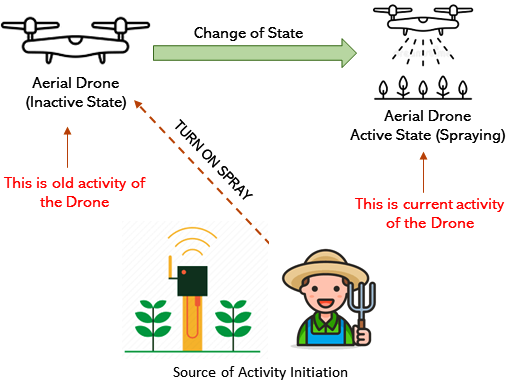}
\centering
\caption{Overview of an Activity.}
\label{fig:activity}
\end{figure}



Figure \ref{fig:activity} provides an overview of an activity (or state) transition from inactive to spraying. This intuitive illustration inspires the following discussion regarding activity primitives.


\textbf{What is an Activity?} An activity is the current state of a device. It signifies what the device is currently performing. A device can perform one activity, or it can perform multiple activities at a time simultaneously. For example, a smart manufacturing machine may have robotic arm for packaging and cooling fan to lower the temperature of the packed product. Similarly a clustered object \cite{gupta2018authorization, gupta2019dynamic}, like a smart car can have multiple sensors within it, and therefore can have multiple activities simultaneously initiated by different sensors inside it. Intuitively an activity is a \textit{long} continuous event occurring for \textit{some} time. It embodies the coarse-grained state of the entity, relevant to making decisions about authorized operations that transition amongst different activities of the entity. The fine-grained state of the entity will typically involve additional parameter. For example, in Figure \ref{fig:activity} the drone in the spraying activity state will have speed, heading, altitude and so on as parameters of its fine-grained state. 

\textbf{What are the entities involved in an activity?}
An activity involves sources who initiate or request the activity, objects on which the activity is requested, and an operation performed by the subject which results in the initiation or change the state of object to start a new activity. In addition, there could be various conditions which must be satisfied to allow an activity. These conditions can impact other activities in the system, or may need a two step approach to get approval from the owner/manager of the device. Further, environmental and contextual factors will also play an important role to permit or deny an activity on an object. Sometimes, an activity may not need a source to initiate an activity, but an event which triggers an activity to happen. For example, if someone breaks the window glass, a sensor will start the \textit{alarming} activity or \textit{vibrating} sensor due to low fuel level in the oil tank. 

\textbf{How an activity is initiated?}
An activity is the result of an operation requested by different subjects or can result from event(s) in the system. The operation requested by a subject or event will result in the start or abort of an activity performed by the smart object. Such subjects can be users in the system or various other objects which can request an operation on other objects to perform an activity. For example, in smart farming ecosystem a soil moisture sensor can request a "TURN ON" operation on \textit{water sprinkler} which \textit{can} result in activation of \textit{sprinkling} activity on water sprinkler provided the moisture sensor has the permissions along with checking that the requested activity itself is allowed based on other conditions in the entire collaborative system. An activity on an object can also be initiated as a result of other activity on same or different related objects. For example, opening the air panes in a greenhouse at 2 am on a snow day will trigger the alarm on farm managers' phone.  



\textbf{Associated conditions, obligations and mutability.}
To allow an activity on a device, there could be pre-, current- or post- conditions which must be satisfied in the connected ecosystem. These conditions can be other activities, system constraints, obligation to start or abort an activity, limits on the number of times an activity can be requested etc. to enable a particular activity. For example, a sibling can turn off the smart speaker of his/her brother if playing after 12 am. 
Also, if a sibling is studying in the room, the smart speaker is not allowed to be turned on, i.e. the current activity of the sibling limits the activity of other siblings. Further, environmental conditions may involve factors, for example, if the humidity level is less that a particular value, then only humidifier can be turned ON by the humidity sensor. In addition, limits on activities can also be supported, for example, pest spraying is only allowed twice a week, irrespective of who requests the activity.


\subsection{Activity Relation Characterization}
In a collaborative environment, different functions and activities work together offering a real automated, smart and efficient CPS. These activities can be disjoint, or mutually exclusive but can also be inter-dependent on each other in particular order, precedence or other relationship characterization. 

In this section, we illustrate how different activities in the system can be characterized. Our discussion  primarily characterizes the relationship using only \textit{two} activities, but this can be extended to any number of activities in the connected ecosystem. Also, these activities can be completely separate or unrelated which can initiated on multiple different or the same device (in case a smart object supports multiple activities, for example, an aerial drone performing thermal imaging and pest spraying). Further, the role or relationship (owner of the object) of activity requesting subject (user or device) with respect to the object on which the activity is requested along with other factors (discussed later), will determine the activity access control decision. The proposed characterization may have overlapping use case scenarios but are important to have separate discussion for each relationship. In this discussion, we will only characterise related activities as shown in Figure \ref{fig:relation} (leaf nodes) and discuss other relevant factors in the subsequent section. 


\begin{enumerate}[label={\arabic*.}]
    \item \textbf{Ordered:} These set of activities are only allowed if initiated in a particular order. In this type of activity relation, if an activity A needs to be started on an object, then either activity B must already be initiated or must be started after activity A is complete  on same or different device. These ordered and interdependent activities can be requested by same or different subjects, which is immaterial. Sequential activities cannot be started out of order, and can also have consequential impact for future activities. For example, in smart farming, \textit{thermal imaging} is activated on the aerial drone by autonomous tractor only after the drone is done with the pest spraying. Or \textit{turning on} water sprinkler is allowed only after tractor has ploughed a farm. Each activity (thermal imaging or water sprinkling) cannot be activated without checking the requisite preceding activities. In this example, the issuing source is irrelevant, but can be important in different scenarios (e.g., farm manager request may not be constrained to be in this order).
    
    \item \textbf{Concurrent:} These set of activities must always happen simultaneously or alternately are allowed to happen simultaneously. This involves activities which are related to each other, or the activities which are completely disjoint and have no impact (relation) among themselves. It is also possible, that activities can be allowed to be concurrent but on different devices, and not on the same device. In this case, if an activity A has started, then activity B must also be initiated in parallel. For example, when the nutrient spraying activity is started, the sensor measuring the nitrogen level in the soil must also be activated so that crops do not receive over supply of nutrients. These activities are allowed to be initiated on the same device also which has the capability to spray as well as to record the nitrogen level. The concurrent activities may be initiated by same subject, or require different subjects to start activities concurrently. In that case, if a subject A tries to start an activity, it may trigger a request or alert to another subject B who should initiate other concurrent activity(s).

    
    
    
    
    \item \textbf{Temporary:} These related activities can be allowed \textit{sometimes} based on the conditions depending on the context (environmental factors) or the acting user/subject (could be the administrator of the business unit). This is applicable to activities which can happen on same or different devices, and possibly on the relationship among the device owner and the acting user. For example, an activity A is allowed with activity B temporarily, only when the issuing user is the admin of the object, and if there is an alert/emergency in the system. In case of a greenhouse, roof ventilation system can be closed by the farm manager while the pesticide spray is happening only during a tornado warning. Otherwise, these two activities can never be allowed to happen at the same time outside the tornado condition.
    
    \begin{figure}[t!]
\centering
\includegraphics[width=\columnwidth]{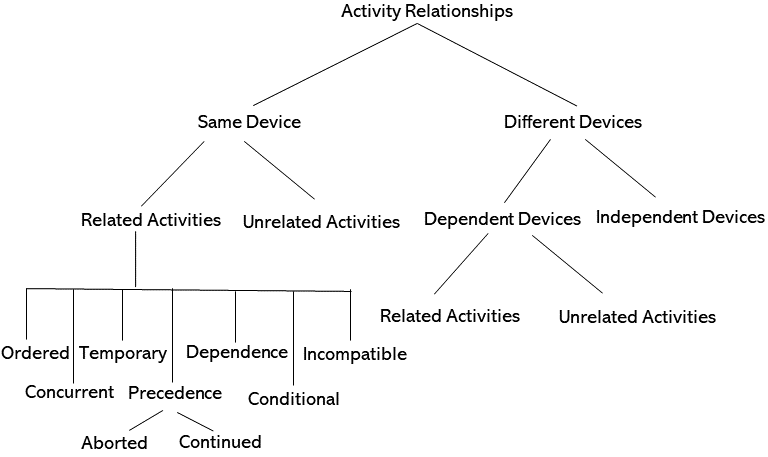}
\centering
\caption{Activity Relation Characterization}
\label{fig:relation}
\end{figure}
    
    
    \item \textbf{Precedence:} There can be an activity which always take precedence over other set of activities. If such an activity is requested, it is possible that some existing (or current) activities in the system are \textit{aborted} (pre-empted or halted temporarily) while some can be continued. It is also possible a new activity may only be allowed for certain amount of time, or if the context is important, and other halted activities can continue after activity is complete. One activity may supersede another, for example, calling 911 by parent using Alexa if fire has been detected in the house. This is always allowed, irrespective of what other activities, for example, playing music, is currently active on Alexa. Another example, if a nutrient solution unit is ON (i.e mixing a new concentration), then all nutrient spraying actuators in the field must be stopped. This may not be the case for actuators located in farm area not supplied by the unit. Spraying actuators can continue after nutrient mixing is complete.

    \item \textbf{Dependence:} These set of activities have dependence on each other. Such dependence can be concurrent, preceding or succeeding. Therefore, concurrent can be considered as a sub-case for these set of activities. In this case, for instance, an activity A should be active to allow requested activity B or it can be case such that activity A will be allowed but activity C will also start either in parallel or after. Similarly, there could be a case when some activities can never be allowed to perform together. For example, the humidifier in the air cooling system can operate only along with thermal shading activity in greenhouse bay unless activated by the greenhouse manager. Similarly, after the pest spray activity is done by the aerial drone, weed scanning activity must always be initiated, and the thermal imaging activity by the cameras should be stopped.




    
    \item \textbf{Conditional:} 
    These related activities can only be allowed if any particular condition (or conditions) is satisfied. These conditions can be related to the location of the objects on which the activities are initiated, or if the requested related activities are on same or different device, or permitting a new activity may require another set of activities to be started or stopped. As an example, an activity A is allowed with activity B, only if they are performed on different devices located in different rooms. Also at the same time activity C must be initiated by the administrator. In smart farming, ventilation system in the greenhouse can only be turned on after the pest spraying activity has been completed, and not before 1 hour after spraying. In another case, a sibling can raise the thermostat temperature to 75 in his brothers' room if he is sleeping alone, and his current location is inside the same room.

    
    
    
    
    
    \item \textbf{Incompatible:} These set of activities can never be allowed to take place simultaneously, or one after the other, or may be not within a time span. This can also be true irrespective of who initiated it, or any contextual emergency. For example, on the same crop field, pesticide spraying by the drone and water spray by the sprinkler cannot be allowed to happen together, or within 2 hours after any of the activities has been completed. Similarly, fogging in the greenhouse and opening windows cannot be performed together. An autonomous tractor cannot be issued operations to plough the crop field and remove weeds at the same time. By and large, these activities can never be allowed together unless special conditions need. 
    
\end{enumerate}
  
As shown in Figure \ref{fig:relation} the relationship between activities can be mainly characterized based on whether they are on the same or on different devices. If a new activity is initiated on a device which is different than the devices on which existing activities are running, and the devices are independent, then there is no conflict for activity to be allowed. If related activities are initiated on dependent devices, access control decision to allow the activity or not should be based on other activities in the system. This branch is similar to the characterization in which activities are requested on same device. Further, in case the activities are unrelated and do not impact each other, it is immaterial if they are on same or different device. Such activities are allowed. As discussed in the previous subsection, various \textit{relations} among activities in the ecosystem control how and which activities are allowed. Therefore, dependence among devices/objects and among the activities is important to make a decision. The independence of the objects can be based on the location, for example if they are on separate farm areas or in different manufacturing units, or it can be functionality of devices. Similarly, some activities may be allowed to be performed together, but on different devices.

These aforementioned characteristics can be combined and have numerous different  combinations to express different security conditions. In addition, there are other factors which will require finer grained policy expression as discussed in the following subsection. It should be noted that this characterization is not exhaustive, and new cases (relationships) will arise and mature as we examine additional use cases in the proposed notion of activity-centric models.

\subsection{Factors Impacting Activity Decision}

The activities allowed or denied in the system can depend on multiple factors including other activities (as discussed in previous subsection), users or devices requesting the activity, environmental factors, dependence among different objects, count on the activity performed over a period of time, and user or device permissions. Some questions which need to be considered to allow a new activity in the system include: a) Where is the activity allowed to perform? For example, a child may be allowed to turn on smart speaker in his room, but may not be allowed to do the same when father (or any other relation) is watching TV in living area; b) Who is allowed to do it? This can be based on the identity of the user or it can also be dependent on the relation between the device owner and the requesting user or device; c) When is the activity allowed? It could be an emergency, high system risk, number of times the activity has been requested in a time frame or anomaly in the system. It should be noted that before any other factor is checked to initiate an activity, it is imperative to ensure that the user or the device has the permission to perform the operation on an object to trigger the activity. If the subject does not have the permission the requested activity should be denied regardless of other circumstances. 


\begin{figure}[t!]
\centering
\includegraphics[width=\columnwidth]{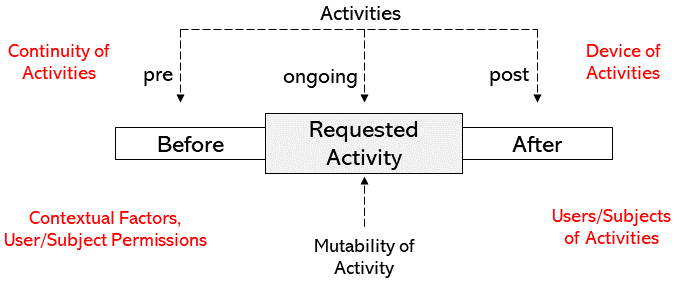}
\centering
\caption{Factors Impacting Activity in Collaborative System}
\label{fig:factor}
\end{figure}

As illustrated in Figure \ref{fig:factor}, there are several other factors which also contribute in the activity decision. Identity, attributes or relationship of \textit{Users} or \textit{Subjects} who initiated or are requesting an activity in the system are among these. A smart farm owner can start the water sprinkler which will result in aborting the pest spray activity initiated by the temporary worker. However, this may not be the case when owner the started the pest spray activity and worker tries to water sprinkle. Therefore, role and attributes of the user and subjects are critical along with the relation among activities. There can be a relation between object and source (user or subjects) or among different objects which may be impacted by the activity request. For example, they can belong to the same owner, or same farm, same production unit, same room, same household etc. 
In addition, similar to UCON \cite{sandhu2003usage}, a user/device may have a \textit{threshold limit} on how many times a particular activity can be initiated. This limit can be per device  which are allowed to activate the activity (drone can pest spray only once or weed detector sensor can initiate pest spray once via drone), or it can be system wide, such as pest spraying is only allowed twice a day irrespective of how it was initiated or who performed the activity. In addition, the `age' of the device in the ecosystem may also increase or decrease its permissions on the set of activities it can perform. A new user or device can be authorized to a limited set of activities. As the user or device gets older (meaning it has been associated with the organization or the functional unit for a longer time), the set of permissions can increase. Continuity of activities will also be impacted with the request for new activities, in that an ongoing activity may be halted or aborted. This property (inspired by UCON) is called ``continuity'' and has to be captured in modern access control for the control of relatively long-lived activity or for immediate revocation of activity. In addition, activities can also be restricted within a time window, for example, activity A cannot be allowed within 4 hours post completion of activity B. 



















\section{Activity Control Expression}
In this section, we sketch a preliminary approach to represent policies controlling activities in the system using activity control expressions. We will provide a generalized structure and representation of expression using different characterization scenarios and access control factors as discussed in the preceding sections. Our attempt is not so much to come up with a formal policy language, but rather to have a sample overview of how such activity-centric policies can be represented. 

A \textbf{generalized} activity control expression defined per smart object in the ecosystem can be specified as follows.

\noindent
Object: \textbf{ObjectX}

\noindent
$\langle$ (op, sourceX, activityX), \\ ($pre\_conditions$), \\ ($current\_conditions$),\\ ($resulting\_conditions$),\\ ($contextual\_conditions$) $\rangle$
where \\
$pre\_conditions$, $current\_conditions$ and  $resulting\_conditions$ can be expressed as $\langle$ ($pre\_state, objA, sourceY$), ......  $\rangle$, \\ $\langle$ ($cur\_state, objB, sourceZ$), ......  $\rangle$, and
as \\ $\langle$ ($new\_state, objC, sourceX$),..$\rangle$ respectively. 

\sloppy
This \textit{generalized} policy is defined for object ObjectX, stating that sourceX is allowed to perform an operation op which will change the state of \textit{ObjectX} to a new activity activityX, iff the $pre\_conditions$ and $current\_conditions$ are satisfied together with the $contextual\_conditions$, and the resulting conditions for the approval of this requested activity activityX is stated in $resulting\_conditions$. These conditions capture the pre- and current conditions for ObjectX as well as other related/dependent objects (\textit{objA, objB, ....}) and their corresponding activities ($pre\_state, cur\_state,new\_state$) as well as the subjects (\textit{sourceY, sourceZ, ....}) which initiated that activity in the collaborative system. In addition, the $contextual\_conditions$ can be defined in first order propositional logic formula with a suitable policy language such as \textit{value(nitrogen)} $>$ 50 $\land$ \textit{value(weather) = severe}. It should be noted that usage control for each activity (the number of times the activity can be performed during a particular time-frame, for example, water spraying can only be performed twice a day irrespective of which subject is requesting) along with per source activity control (a subject may only be allowed to activate an activity only three times in a week) must also be expressed in activity control policy expression. 
Following are some examples of using the generalized activity control expressions in different  scenarios. 
\subsection{Activity Control Policies in Smart Farming}

\noindent
\textit{\underline{Activity Example 1:}}  A soil moisture sensor can issue operation TURN ON to a water sprinkler and change its state to active (i.e. water sprinkler is running), only if the water sprinkler is currently in an inactive state which was changed by farm-manager (this is a pre-condition).

\noindent
\textit{Object}: \textbf{ Water Sprinkler}

\noindent
$\langle$ (TURN-ON, moisture sensor, Spraying), \\ 
(cur\_inactive, Water Sprinkler, farm-manager) $\rangle$.  

\noindent
\textit{\underline{Activity Example 2:}}  A weed detector sensor can issue operation SPRAY ON to an Aerial Drone and change its state to Spraying, if the Nitrogen sensor in the field is active (i.e sensing nitrogen level in soil) enabled by farm-manager and the nitrogen level in soil is less than 50.

\noindent
\textit{Object}: \textbf{Aerial Drone}

\noindent
$\langle$ (SPRAY-ON, weed detector, Spraying), \\ 
(cur\_active, Nitrogen sensor, farm-manager),\\
(value(nitrogen-level) $<$ 50) $\rangle$.

\noindent
\textit{\underline{Activity Example 3:}}  Thermal imaging is activated on the Aerial Drone by autonomous tractor only after the drone is done with the spray which was initiated by weed detector, and as a consequence of the start of the new activity (i.e Thermal-imaging) the spraying should be stopped.

\noindent
\textit{Object}: \textbf{Aerial Drone}

\noindent
$\langle$ (IMAGING-ON, autonomous tractor, Thermal imaging), \\ 
(cur\_spraying, Aerial Drone, weed detector),\\
(new\_inactive-spraying, Aerial Drone, autonomous tractor) $\rangle$. 

\noindent
Here, it should be noted that the resulting activity i.e. spraying to stop is due to the new activity (Thermal imaging) requested by autonomous tractor. Therefore, autonomous tractor is also the source for inactive spraying on Aerial Drone object.

\noindent
\textit{\underline{Activity Example 4:}} Turning ON Pest Spray after Tractor has ploughed a farm (assuming worker issued command for the tractor to plough) is allowed irrespective of source of the activity request.

\noindent
\textit{Object}: \textbf{Pest Spray}

\noindent
$\langle$ (TURN-ON, ANY, Spraying ), \\ 
(pre\_ploughing, Tractor, worker) $\rangle$. 

\subsection{Activity Control Policies in Smart Manufacturing and Industrial IoT}

\noindent
\textit{\underline{Activity Example 5:}}  The filtering process can  be started on smart Oil Filter by the production manager, if the oil tank valve is closed.

\noindent
\textit{Object}: \textbf{Oil Filter}

\noindent
$\langle$ (TURN-ON, production manager, Filtering), \\ 
(cur\_close, Oil tank Valve, ANY) $\rangle$.

\noindent
\textit{\underline{Activity Example 6:}}  Air Conditioner cooling in the pharmaceutical facility cannot be turned on by the thermostat if the moisture sensor is inactive, and the current temperature is greater than 75.

\noindent
\textit{Object}: \textbf{Air Conditioner}

\noindent
$\langle$ (TURN-ON, thermostat, Cooling), \\ 
(cur\_active, moisture sensor, ANY), \\
(value(temperature) $>$ 75)$\rangle$.

\noindent
\textit{\underline{Activity Example 7:}} Oil pumping by the smart valve can be activated only after hydrotreating is performed by the hydrotreater unit by worker. After oil pumping, outlet valve should remain closed.

\noindent
\textit{Object}: \textbf{Tank Pump}

\noindent
$\langle$ (TURN-ON, valve, pumping), \\ 
(pre\_hydrotreating, hydrotreater-unit, worker), \\
(post\_closed, outlet valve, ANY)
$\rangle$.

\noindent
\textit{\underline{Activity Example 8:}}  Robotic Arm must be inactivated when production belt accelerometer is vibrating.

\noindent
\textit{Object}: \textbf{Robotic Arm}

\noindent
$\langle$ (Inactive, ANY, Inactive), \\ 
(cur\_vibrating, production belt, ANY) $\rangle$. 

\noindent
Here, ANY is used when the sensor does not need a subject to initiate an activity. Here the belt started vibrating due to some event. Therefore activity can be initiated by subjects or an event. 

\subsection{Activity Control Policies in Smart Home}

\noindent
\textit{\underline{Activity Example 9:}}  A user with mobile phone can only be allowed to unlock the door alarm if parent has inactivated the alarm.

\noindent
\textit{Object}: \textbf{Door}

\noindent
$\langle$ (UNLOCK, mobileX, Inactive), \\ 
(cur\_inactive, alarm, parent) $\rangle$.  

\noindent
\textit{\underline{Activity Example 10:}}  Bob can turn off the smart speaker of Tom if playing after 12 am.

\noindent
\textit{Object}: \textbf{Speaker}

\noindent
$\langle$ (TURN-OFF, Bob, Inactive), \\ 
(Active, Speaker, Tom) , \\
(value(Time) $>$ 12) $\rangle$.

\noindent
\textit{\underline{Activity Example 11:}}  A child may be allowed to turn on smart speaker in his room, but may not be allowed to do the same when parent is watching TV in living area (this is current condition).

\noindent
\textit{Object}: \textbf{Speaker}

\noindent
$\langle$ (TURN-ON, Child, Active), \\ 
(cur\_inactive watching, TV, Parent) $\lor$ \\
(cur\_active watching, TV, Parent) $\land$ location(TV) != living area $\rangle$.

The aforementioned policies represented are defined per object, which is not the optimal way to design such a policy based solution which may have hundreds of smart objects in a connected ecosystem to be controlled. One approach to define these policy is per \textit{object type}, which is similar to an attribute assigned to objects which have alike functionality or perform same activity. We can also separate objects into \textit{object groups}, so as to have similar objects into one category and define a policy per group. Such number of groups will be less, easier to administer and define policy per group. Similar groups can be created for subjects or source of the operation who requested the activity. A sample policy per object and source groups may be represented as follows. 

\noindent
\textit{\underline{Activity Example 12:}}  Window in a greenhouse cannot be opened by any user or sensor during the pest spraying activity initiated by any user on any device. (Here Window is the object type, or group of all windows in the greenhouse.)

\noindent
\textit{Object-Type}: \textbf{Window}

\noindent
$\langle$ (TURN-ON, ANY, Open), \\ 
(cur\_inactive Pest Spraying, ANY, ANY) $\rangle$.

In this section, we illustrated our first attempt in an informal manner to create and design activity control expressions with examples in different connected and collaborative domains. Additional research is needed to express these policies in a more understandable and structured manner. 




\section{Future Research Agenda}
\begin{figure}[t!]
\centering
\includegraphics[width=\columnwidth]{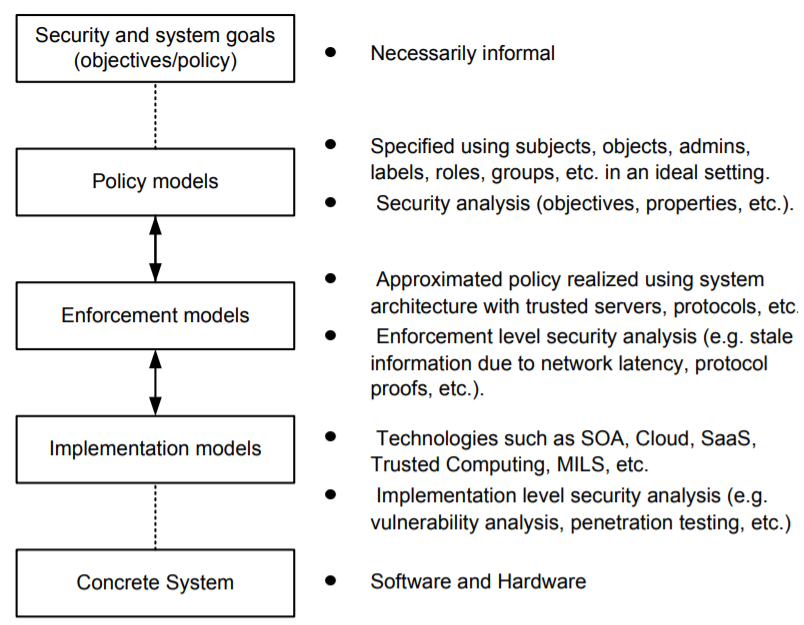}
\centering
\caption{PEI Models Framework \cite{sandhu2009pei}}
\label{fig:pei}
\end{figure}

So far we have proposed a novel perspective and approach to address access control needs in dynamic and activity oriented collaborative ecosystem considering \textit{activities} as the fundamental entity. We discussed activity characterization, and highlighted core requirements to define activity-centric access control (ACAC) system focusing on the conceptual foundational principles in an informal way. In this section, we highlight future research directions needed to fully mature activity-centric models in connected CPS. We will use the PEI framework \cite{sandhu2009pei} as illustrated in Figure \ref{fig:pei} to highlight the research requirements. These are the Policy (P), Enforcement (E) and
Implementation (I) layers, and formal models are needed to express and analyze the security policy at each of these three layers. Following we highlight some open research questions. 

\subsection{Access Control Operational Model and Extensions} 

Formal mathematically grounded models and outline abstractions similar to RBAC96 \cite{sandhu1998role} and ABAC \cite{jin2012unified} are needed to define different entities included in proposed ACAC request decision. It is desirable to define a meta-model which provides the foundation for designing extensible and adaptable access control models which can fit into different CPS domains. These models should address the core issues of pre, current and post activities along with condition, obligations, mutability and usage needs per user or activities. Subjects, objects, and operations 
can be divided into several detailed components with different perspectives. In addition, the foundational models must address the principles of next generation access control \cite{bhattconvergent}. This reflects the need for security models at layer 1 and 2 in the PEI model shown in Figure \ref{fig:pei}.  The main  motivation for developing formal models is to focus on the real policy needs
of the application without being distracted by implementation
details and practical realities of smart connected systems.

\subsection{Enforcement Architectures}
It is important to identify different cloud, edge, or hybrid architectures to enable deployment of the formal models. Such architecture will be formulated based on the application needs. Further, federated and collaborative architectures also need to be developed to enable cross enterprises, cross domain interaction and communication among smart devices and controlling their activities in shared connected ecosystems such as co-operatives. This will also deal with the approximations and additional servers
introduced by the distributed nature of real-world distributed
systems. The goal should be to make the approximations explicit and
controllable since perfect correspondence to the idealized models
layer is impossible. Different architectures are possible based on where the policies are maintained, evaluated, and enforced. Enforcement is key issue and is particularly critical for IoT devices which are resource constrained.  

\subsection{Policy Language and Constraints}
Policy and activity control expression languages must be developed expressing the policies specifiable by the formal operational and administrative models developed. In addition, the policies must be flexible and extensible to apply in different domains. These policies must consider not only the subjects, objects and activities, but also consider conditions and obligations along with usage control to address the mutability aspect of an activity. The policy language supporting the activity control expression must be able to capture if the activity is not requested beyond a numerical threshold, or by the threshold per subject. Theory must be developed for constraints specifications such as separation of duty and cardinality constraints regarding contextual risk factors. An initial approach would be develop a framework to first understand different kinds of constraints. Such policy language is also important to represent multiple relationships and characterization among different activities as discussed in Section 3. 

\subsection{Administrative Models}
Administrative models provide facilities to create entities, and define conditions, contextual factors, activities and correlations, relationships etc. in the system. Such models enable specification of who can decide on devices and activities which impact different sections and working units in a single large domain. For example, a device in the greenhouse of the farm may be impacted by the food processing unit in the same farm, but these are managed by two different security administrators. So, how will the activities be checked if they impact objects in these two sub-units. Can the owner of the object, or the manager of the location of the object make such decision? Who will allow to authorize the activity? Devices can be administered by the owner of the device, but can also delegate administration to someone else after initial setup. For example, a farm owner can temporarily delegate activity access rights to temporary workers. It is important to examine if ABAC and RBAC like administrative models can work in the activity focused domains, or an activity based approach is needed to develop similar models for proposed ACAC models. These questions becomes more interesting in distributed domains administered by different users, where multiple activities are happening concurrently. 

\subsection{Convergence with Access Control Models}
Bhatt and Sandhu \cite{bhattconvergent} proposed access control convergence reflecting the need for \textit{crossbreed} or \textit{hybrid} of established and new access control models to support dynamic and distributed future connected systems.  In such domains, attributes of different entities, relationships between different entities, and other features must be captured to provide
dynamic and fine-grained access control that can be adapted and enforced in a wide range of applications. Similar convergence is also needed for activity-centric models. Since the activities provide a two step check, one if the subject is allowed to perform an operation to start an activity, and second, if there are activities, constraints or obligations among activities is satisfied, to allow or deny an activity, such convergence is natural. The goal should be to develop and evaluate such hybrid approaches, and determine their applicability and expressiveness to support policies among different distributed applications and CPS domains. 

\subsection{AI and Data Driven Deployment}
AI aims to support automated security defense solutions. Similarly research is needed to develop AI and data driven systems based on activity logs and other important data points to automatically define the relationship among different activities in the system, and develop a self-adaptive AI based ACAC mechanisms. Data from the connected ecosystem can be used to define the activity-centric policies, or AI models can be trained to make a decision without the need for explicit policies as such. However, how to ensure if the AI models will work as intended or decide correctly is also a research question. Attempts have been made to develop data driven system \cite{salapura2019generative, jabal2020polisma} which allow to generate optimal access control policies for autonomous systems. Similar frameworks are needed for activity-centric access controlled CPS systems. This supports research at the implementation layer in Figure 4. 

\subsection{Safety Analysis}
It is important to conduct the theoretical analysis of both administrative and operational activity-centric models that will be developed. A primary safety question is whether, a system can reach to a state when certain activities be allowed by subjects under certain conditions. Because some of the smart domains (such as water treatment facility) are very critical, it is imperative to ensure that activities relationships, along with pre or post conditions must always be satisfied. A sample safety question can be, is it possible that system reaches to a state where two conflicting activities are allowed by different subjects, for example, pest spray and water spray in a smart farm.





\subsection{Application in CPS domains}
The developed models and mechanisms for the proposed activity-centric access control must be adaptive to be applied in different smart and connected domains. In our activity control expression, we have attempted to reflect some activities in smart farming, manufacturing and homes. We firmly believe that other CPS domains such as transportation \cite{gupta2020attribute}, building, healthcare, energy, defense, robotics etc. will have similar concept of interconnected activities along with notion of obligation and conditions among different smart entities. Once a core formal model is developed, these should be extended to various smart, distributed and collaborative systems.

\section{Conclusion}
Activities are an integral part of a connected and coordinated CPS.  Different sensors, actuators and smart devices perform various activities in the system which collectively result in the automation of an entire ecosystem. These fully integrated and collaborative systems respond in real time to meet challenges of growing distributed but synergistic smart systems supported by IoT. This paper outlines our first attempt towards a vision of activity-centric access control framework for smart connected ecosystems. The core idea is to how the current or preceding activities along with other important factors including usage, conditions and obligations for activities limit access control for new activities requested by users or devices on different smart objects. We discuss and characterize relations among various activities. In addition, a preliminary attempt is made to define activity control expressions in different smart use-cases which offers a structure considering pre, current and post activities along with contextual conditions to control activities. We outline a future research agenda envisioning development of formal models and policy languages along with enforcement architectures for the proposed activity-centric access control solutions. 

\section*{Acknowledgement}
This research is partially supported by NSF CREST Center Grant HRD-1736209 at UTSA, and by the NSF Grant 2025682, Faculty Research Grant Program and NASA Grant 80NSSC20M0186 at Tennessee
Technological University.


\bibliographystyle{unsrt}
\bibliography{sample-base}

\end{document}